\documentclass[twocolumn,english,aps,prl,showpacs,groupedaddress]{revtex4}
\usepackage{subfigure}
\usepackage{graphics}

\usepackage{graphicx}
\usepackage[T1]{fontenc}
\usepackage[latin9]{inputenc}
\usepackage{ifsym}
\usepackage{wasysym}
\usepackage{babel}
\usepackage{epstopdf}
\usepackage{color}
\usepackage{soul}

\begin{document}

\title{Drag enhancement and drag reduction in viscoelastic  flow}

\author{Atul Varshney$^{1,2}$ and Victor Steinberg$^{1,3}$}
\affiliation{$^1$Department of Physics of Complex Systems, Weizmann
Institute of Science, Rehovot 76100, Israel\\$^2$Institute of Science and Technology Austria, Am Campus 1, 3400 Klosterneuburg, Austria\\$^3$The Racah Institute of Physics, Hebrew University of Jerusalem, Jerusalem 91904, Israel}


\begin{abstract}
 Creeping flow of polymeric fluid without inertia exhibits elastic instabilities and elastic turbulence accompanied by drag enhancement due to elastic stress produced by flow-stretched polymers. However, in inertia-dominated flow at high $\mbox{Re}$ and low fluid elasticity $El$, a reduction in turbulent frictional drag  is caused by an intricate competition between inertial and elastic stresses.  Here, we explore the effect of inertia on the stability of viscoelastic flow in a broad range of control parameters $El$ and $(\mbox{Re}, \mbox{Wi})$. We present the stability diagram of observed flow regimes in $\mbox{Wi}-\mbox{Re}$ coordinates and find that instabilities' onsets show unexpectedly non-monotonic dependence on $El$. Further, three distinct regions in the diagram are identified based on $El$. Strikingly, for high elasticity fluids we discover a complete relaminarization of flow at Reynolds number of the order of unity, different from a well-known turbulent drag reduction. These counterintuitive effects may be explained by a finite polymer extensibility and a suppression of vorticity at high $\mbox{Wi}$.  Our results call for further theoretical and numerical development to uncover the role of inertial effect on elastic turbulence in a viscoelastic flow.

\end{abstract}

\maketitle
 \section{Introduction}

Long chain polymer molecules in Newtonian fluid alters the rheological properties of the fluid; the relation between stress and strain becomes nonlinear. Moreover, polymers being stretched by velocity gradient in  shear flow engender elastic stress that modifies the flow via  feedback mechanism.  It results in pure elastic instabilities \cite{larson,shaqfeh} and elastic turbulence (ET) \cite{groisman}, observed at $\mbox{Re}\ll1$ and $\mbox{Wi}\gg1$. Here $\mbox{Re}$ is the ratio of inertial to viscous stresses, $\mbox{Re}=UD\rho/\eta$, and $\mbox{Wi}$ defines the degree of polymer stretching $\mbox{Wi}=\lambda U/D$; where $U$ is the flow speed, $D$ is the characteristic length scale, $\lambda$ is the longest polymer relaxation time, and $\rho$ and $\eta$ are  the density and dynamic viscosity of the fluid, respectively \cite{bird}.

ET is a spatially smooth, random in time chaotic flow, which statistical, mean and spectral properties are characterized experimentally \cite{groisman,groisman1,groisman2,teo1,jun,jun1,arratia,atul1}, theoretically \cite{lebedev,lebedev1} and numerically \cite{boffetta,berti,berti1,gotoh}.
The  hallmark of ET is a steep power-law decay of velocity power spectrum with  an exponent $|\alpha|>3$  indicating that only a few modes are relevant to flow dynamics \cite{groisman,groisman1,lebedev,lebedev1}. Further, an injection of polymers into a turbulent flow of Newtonian fluid at $\mbox{Re}\gg1$ reduces the drag and also has a dramatic effect on the turbulent flow structures \cite{toms}. In recent investigations, a new state of small-scale turbulence associated with maximum drag reduction asymptote is observed in a pipe flow at $\mbox{Re}\gg1$ and $\mbox{Wi}\gg1$. This state is termed as elasto-inertial turbulence (EIT) and exhibits properties similar to ET despite the fact that it is driven by both inertial and elastic stresses, and their interplay defines EIT properties \cite{hof,dubief,dubief1}.  Thus, the  fundamental question arises how  inertial effect modifies ET in viscoelastic flow towards turbulent drag reduction.

Numerous studies were performed in various flow geometries to unravel the role of inertia on the stability of viscoelastic flow, albeit contradictory results were obtained. In Couette-Taylor flow between two cylinders, the instability sets in at $\mbox{Wi}_{tr}$ which grows with elasticity number $El$(=$\mbox{Wi}/\mbox{Re}$) saturating at sufficiently high $El$, and reduces with increasing inertia \cite{groismanprl96,groismanepl,groismanphfl}.  Whereas the onset of the instability $\mbox{Re}_{tr}$ is almost constant at very low $El$ and reduces with increasing $El$, in a rather limited range, in agreement with numerical simulations \cite{shaqfeh1,thomas}. Recent experiments in the Couette-Taylor flow with both inner and outer co- and counter-rotating cylinders at low $El$ show weak, smooth dependence on $El$ \cite{muller09,muller11,muller13}. At the moderate $El$, either stabilization or destabilization of the first bifurcation depending on co- or counter-rotation of cylinders is found \cite{muller09}. However, the general tendency in the dependencies of the bifurcations on $El$ reported in Refs. \cite{groismanprl96,groismanepl,groismanphfl} is  confirmed later in Refs. \cite{muller09,muller11,muller13}. On the other hand, a non-monotonic dependence of the first bifurcation in a wall-bounded, plane Poiseuille flow on $El$ in its narrow range of low values is revealed in numerical simulations using the Oldroyd-B constitutive equation. The reduced solvent viscosity strongly modifies this effect: the smaller polymer contribution to the viscosity, the less pronounced effect \cite{suresh}. In extensional viscoelastic flow \cite{mckinley}, e.g. planar flow with an abrupt contraction-expansion, and in  flow past a cylinder  \cite{kenney}, a role of both elasticity and inertia was investigated in a narrow range of $\mbox{Re}$ and $\mbox{Wi}$,  and for only three $El$ values. In extensional flow, the onset of the elastic instability $\mbox{Wi}_{tr}$ turns out to be independent of $\mbox{Re}$ in the range of 0.1 to 40 for three polymer solutions that correspond to $El=3.8$, 8.4 and 89. However, in the case of flow past a cylinder \cite{kenney} $\mbox{Wi}_{tr}$ decreases with increasing $\mbox{Re}$. Recent numerical studies \cite{kellay} on two-dimensional viscoelastic flow past a cylinder reveal the phase diagram in ($\mbox{Wi},\mbox{Re}$) coordinates and both drag enhancement and drag reduction (DR) were observed in the range of $\mbox{Wi}\approx0.2$ to 10 and $\mbox{Re}\approx0.1$ to $10^5$. Thus, despite  extensive theoretical and experimental efforts, the influence of inertia on viscoelastic flow in a broad range of ($\mbox{Re},\mbox{Wi}$) and $El$ is still not understood and a stability diagram of different flow regimes  is missing.

Here we perform experiments, over a broad range of ($\mbox{Re},\mbox{Wi}$) and $El$, in a channel flow of dilute polymer solution hindered by two-widely spaced obstacles  (see  Fig. \ref{fig:expt} for experimental setup). Changing solvent viscosity $\eta_s$ by two orders of magnitude allows to vary elasticity number $El=\lambda(\eta_s)\eta_s/\rho D^2\sim\eta_s^2/\rho D^2$ \cite{liu3} by more than four orders of magnitude. Such approach enables us to investigate the role of inertia in viscoelastic flow in different flow regimes in a wide range of ($\mbox{Re},\mbox{Wi}$) and $El$.

The main feature of viscoelastic flow at $Re\ll1$ between two widely-spaced obstacles is an elastic wake instability in the form of a quasi-2D counter-rotating elongated vortices generated by a reversed flow \cite{atul}. The two vortices constitute two mixing layers with a non-uniform shear velocity profile filling the inter-obstacle space. Further increase of $\mbox{Wi}$ leads to chaotic dynamics with properties similar to ET \cite{atul1}. There are several reasons for the choice of the flow geometry: (i) Since blockage ratio $D/\mbox{w}\ll1$, the flow between the cylinders is unbounded, like "an island in a sea" of otherwise laminar channel flow,  contrary to all previous wall-dominated flow geometries that were used to study ET ($D$ and $\mbox{w}$ are the cylinders' diameter and channel width, respectively) \cite{groisman,groisman1,groisman2,teo1,jun,jun1}. Therefore, it is expected to observe mostly homogeneous, though anisotropic flow, closer to that considered in theory \cite{lebedev,lebedev1} and numerical simulations \cite{boffetta,berti,berti1}. By employing unbounded flow we concentrate on variation in the bulk flow structures due to polymer additives, which results in a significant frictional loss. (ii) Large $\mbox{Wi}$ and $\mbox{Re}$ can be reached in the same system to scan the range from ET to DR. (iii) Several  techniques can be simultaneously employed to quantitatively characterize the flow.

\section{Experimental setup}
 \subsection{The experimental setup and materials}

\begin{figure}[htbp]
	\begin{center}
		\includegraphics[width=8.5cm]{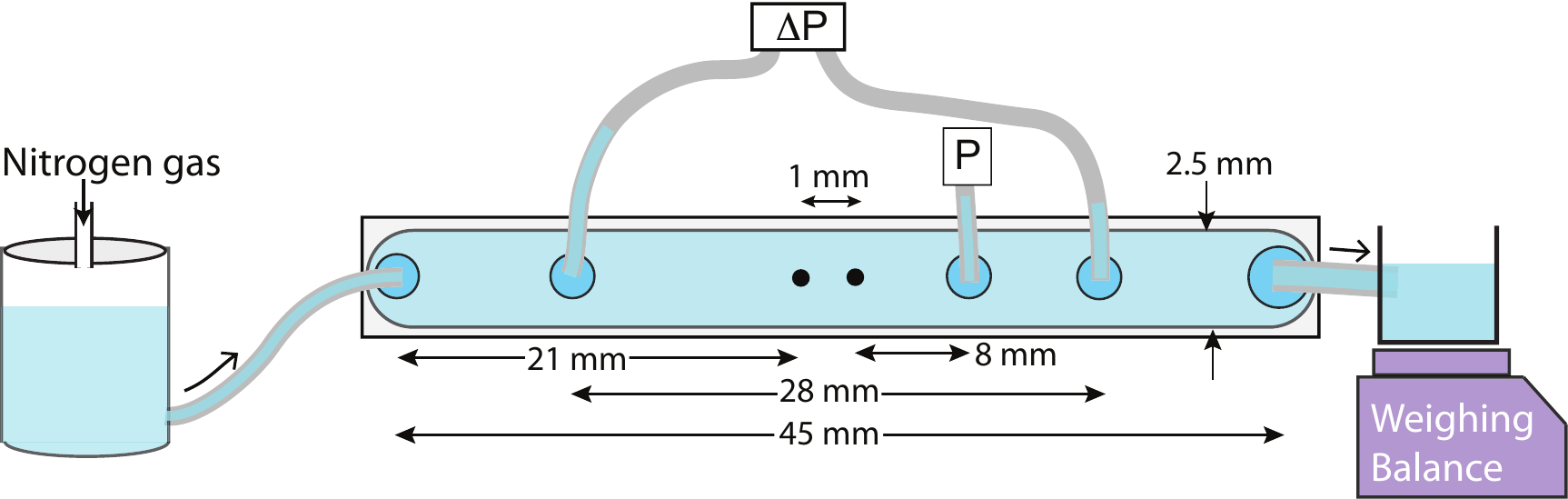}
		\caption{Schematic of the experimental setup (not to scale).  A differential pressure sensor, marked as $\Delta P$, is used to measure pressure drop across the obstacles. An absolute pressure sensor, marked as $P$, after the downstream cylinder is employed to obtain pressure fluctuations.}
		\label{fig:expt}
		\end{center}
\end{figure}

The experiments are conducted in a linear channel of $L\times \mbox{w}\times \mbox{h}=45\times2.5\times 1$  $mm^3$, shown schematically in  Fig. \ref{fig:expt}. The fluid flow is hindered by two cylindrical obstacles of diameter $D=0.30$ mm made of stainless steel separated by a distance of $e=1$ mm and embedded at the center of the channel. Thus the geometrical parameters of the device are $D/\mbox{w}=0.12$,  $\mbox{h}/\mbox{w}=0.4$ and $e/D=3.3$. The channel is made from transparent acrylic glass (PMMA). The fluid is driven by $N_2$ gas at a pressure up to $\sim 60~psi$ and injected via the inlet into a rectangular channel. As a fluid, a dilute polymer solution of high molecular weight polyacrylamide (PAAm, homopolymer of molecular weight $M_w=18$ MDa; Polysciences) at a concentration $c=80$ ppm ($c/c^*\simeq0.4$, where $c^*=200$ ppm  is the overlap concentration for the polymer used \cite{liu3}) is prepared using  water-sucrose solvent with sucrose weight fraction varied from $0$  to $60\%$ (see Table 1 in \cite{sm}). The solvent viscosity, $\eta_s$, at $20^{\circ}\mbox{C}$ is measured in a commercial rheometer (AR-1000; TA Instruments). An addition of polymer to the solvent increases the solution viscosity ($\eta$) of about $30\%$. The stress-relaxation method \cite{liu3} is employed to obtain $\lambda$; for  $\eta_s=0.1~Pa\cdot s$ solution, $\lambda$ is measured to be $10\pm 0.5$ s. Linear dependence of $\lambda$ on $\eta$ was shown in Ref. \cite{liu3}.

\subsection{Pressure measurements and Imaging system}

High sensitivity differential pressure sensors (HSC series, Honeywell) of different ranges are used to measure the pressure drop $\Delta P$ across the obstacles and an additional absolute pressure sensor (ABP series, Honeywell) of different ranges are used to measure the pressure $P$ fluctuations after the downstream cylinder at a sampling rate of 200 Hz, as shown schematically in  Fig. \ref{fig:expt}. The accuracy of the pressure sensors used is $\pm0.25\%$ full scale. We measure pressure drop  both for solvent and polymer solution as a function of flow speed, and the difference between these two measurements provides an  information about the influence of polymers on  the frictional drag.

The fluid exiting the channel outlet is weighed instantaneously $W(t)$ as a function of time $t$ by a PC-interfaced balance (BA210S, Sartorius) with a sampling rate of $5~\mbox{Hz}$ and a resolution of $0.1~\mbox{mg}$. The time-averaged fluid discharge rate $\bar{Q}$ is estimated as $\overline{\Delta W/\Delta t}$. Thus the flow speed is calculated as $U=\bar{Q}/\rho\mbox{w}\mbox{h}$. For flow visualization, the solution is seeded with fluorescent particles of diameter $1~\mu m$ (Fluoresbrite YG, Polysciences). The region between the obstacles is imaged in the mid-plane {\it{via}} a microscope (Olympus IX70), illuminated uniformly with LED (Luxeon Rebel) at $447.5~\mbox{nm}$ wavelength, and two CCD cameras attached to the microscope: (i) GX1920 Prosilica with a spatial resolution $1936\times 1456$ pixels at a rate of $50~fps$ and (ii) a high resolution CCD camera XIMEA MC124CG with a spatial resolution $4112\times 3008$ pixels at a rate of $1~fps$, are used to record particles' streak.

\section{Results}
\begin{figure}[htbp]
	\begin{center}
		\includegraphics[width=8.5cm]{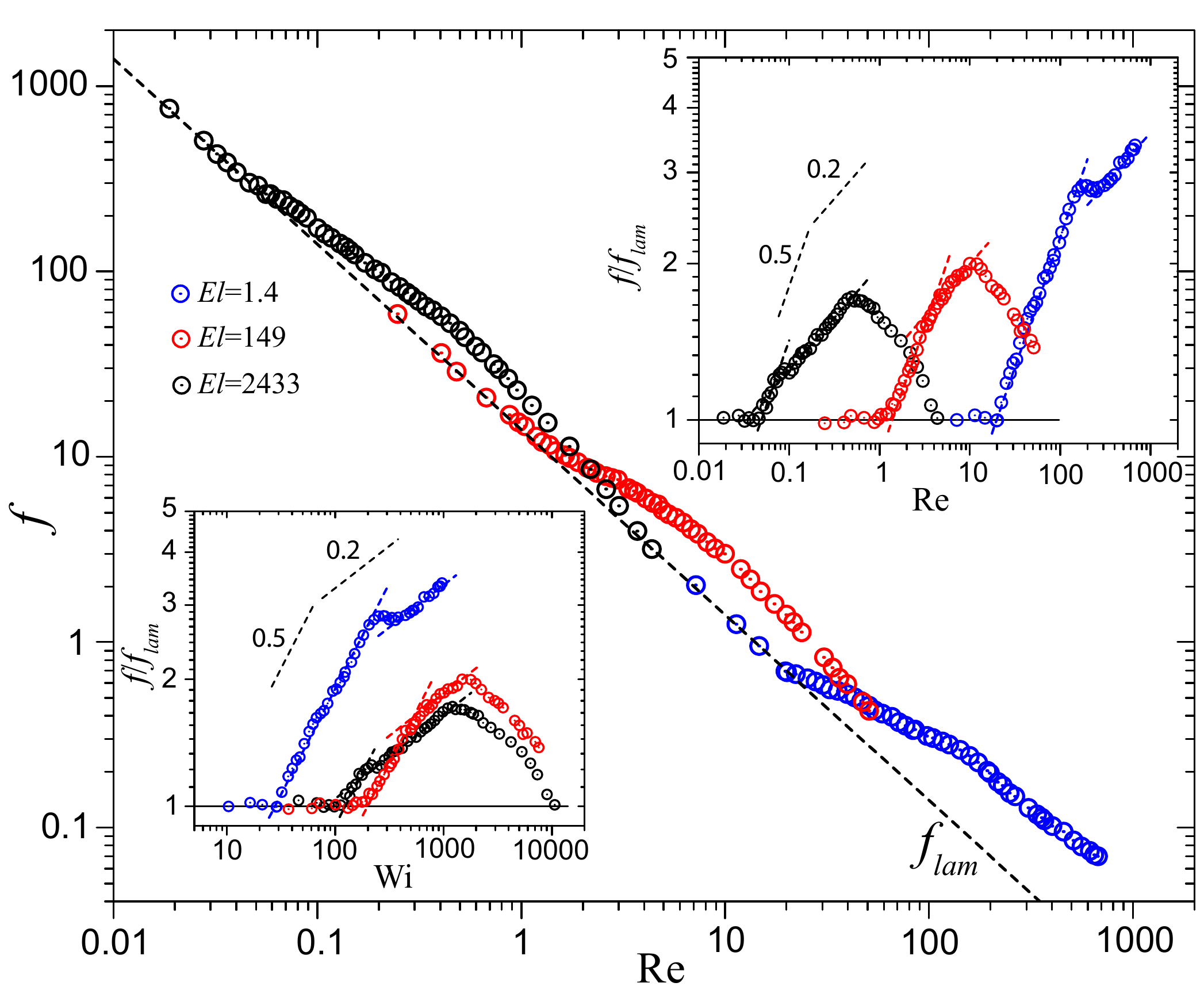}
		\caption{ Friction factor $f$ versus $\mbox{Re}$ for three values of $El$. Dashed line $f_{lam}\sim1/\mbox{Re}$ represents the laminar flow.  Top inset: Normalized friction factor $f/f_{lam}$ versus $\mbox{Re}$. Bottom inset: The same data presented as $f/f_{lam}$ versus $\mbox{Wi}$ with the fits marked by dashed lines in two regions: $f/f_{lam}\sim \mbox{Wi}^{0.5}$ above the elastic instability and $f/f_{lam}\sim \mbox{Wi}^{0.2}$ in the ET regime. One notices the drag reduction for $El=2433$ occurs at $\mbox{Re}\approx0.5$ and $\mbox{Wi}\approx1216$, and continues till the flow relaminarizes.}
		\label{fig:dragcoeff}
	\end{center}
\end{figure}

Frictional drag ($f$) for each $El$ is calculated through the measurement of pressure drop across the obstacles ($\Delta P$) (see Fig. \ref{fig:expt}) as a function of $U$ and is defined as $f=2D_h\Delta P/\rho U^2L_c$;   $D_h=2\mbox{wh}/(\mbox{w}+\mbox{h})=1.43~\mbox{mm}$ is the hydraulic radius and $L_c=28~\mbox{mm}$ is the distance between locations of $\Delta P$  measurement \cite{atul}. Figure \ref{fig:dragcoeff} shows variation of $f$ with $\mbox{Re}$ for three $El$ values and a sequence of transitions can be identified for each $El$. These transitions are further illustrated through  high resolution plot of  the normalized friction factor $f/f_{lam}$ versus $\mbox{Re}$ and $\mbox{Wi}$ presented in  top and bottom insets of Fig. \ref{fig:dragcoeff}, respectively.
Three flow regimes characterized by different scaling exponents are identified: (i) the first drag enhancement  above the elastic instability follows $f/f_{lam}\sim \mbox{Wi}^{0.5}$ for all values of $El$ explored; for high $El$ it is associated with a growth of two elongated vortices (or two mixing layers) \cite{atul}; (ii) further drag enhancement at high $El$ occurs due to  ET \cite{atul1} characterized by a steep algebraic decay in both the power spectra of velocity and pressure fluctuations with the exponents $>\sim3$ (see further), and intensive vorticity dynamics and a growth of average vorticity as $\bar{\omega}\sim\mbox{Wi}^{0.2}$ and $f/f_{lam}\sim \mbox{Wi}^{0.2}$ $-$ typical for ET \cite{atul1}.  For low $El$, either a saturation or reduction of the friction factor with $\mbox{Re}$ or $\mbox{Wi}$ marks the DR regime. And (iii) for both high and intermediate $El$, the DR regime with decreasing $f/f_{lam}$ at increasing $\mbox{Re}$ or $\mbox{Wi}$ is observed and at low $El$, the drag enhancement is noticed. Another striking finding is a complete relaminarization of flow, i.e. 100$\%$ drag reduction,  that occurs for $El=2433$ (also for $El=1070$ and 3704; data not shown), where $f/f_{lam}$ returns back to the laminar value at $\mbox{Re}\approx 4$ ($\mbox{Wi}\approx 10^4$). With decreasing $El$,  the transition points are shifted to a higher value of $\mbox{Re}$ and $\mbox{Wi}$,  and remarkably even at $\mbox{Re}\gg 1$ both drag enhancement and DR regimes can be recognized.

\begin{figure*}
	\begin{center}
		\includegraphics[width=\textwidth]{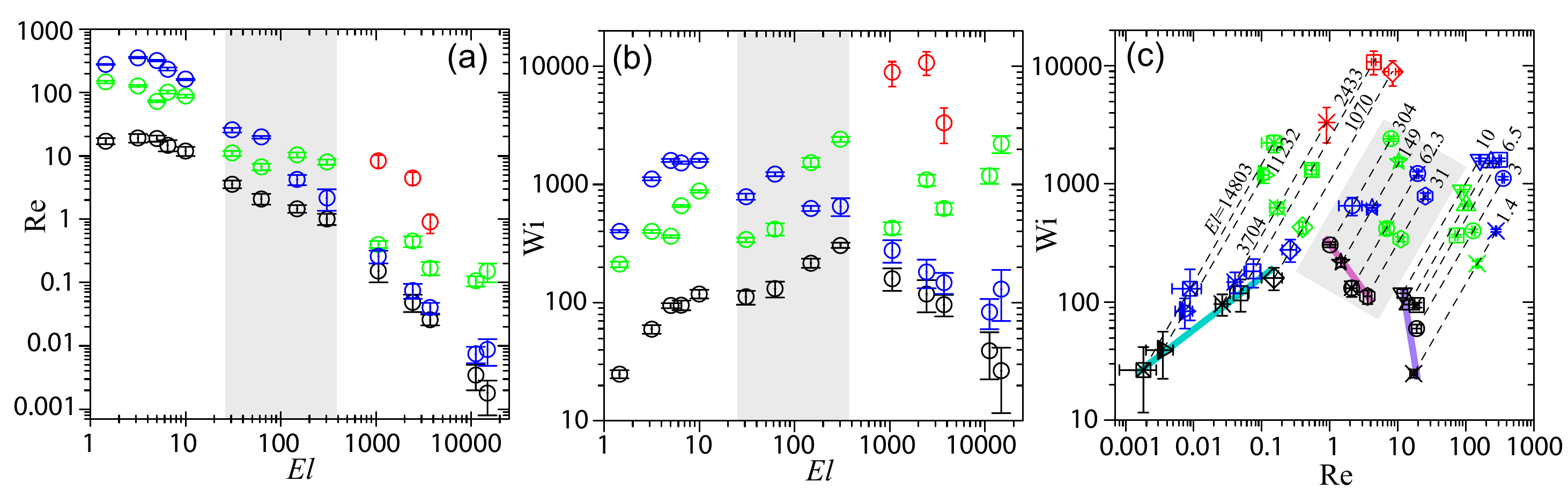}
		\caption{Stability diagram of different flow regimes in (a) $\mbox{Re}-El$, (b) $\mbox{Wi}-El$ and (c) $\mbox{Wi}-\mbox{Re}$ coordinates. The color symbols signify different transitions: first elastic instability (black), DR (green), drag enhancement (blue) and flow relaminarization (red). Grey band in (a-c) indicates the region of intermediate $El$. Solid lines of different colors in (c) are used as a guide to eye to track the transition in different regions.  }
		\label{fig:ReWivsEldiagram}
	\end{center}
\end{figure*}
 \begin{figure*}
	\begin{center}
		\includegraphics[width=17.0cm]{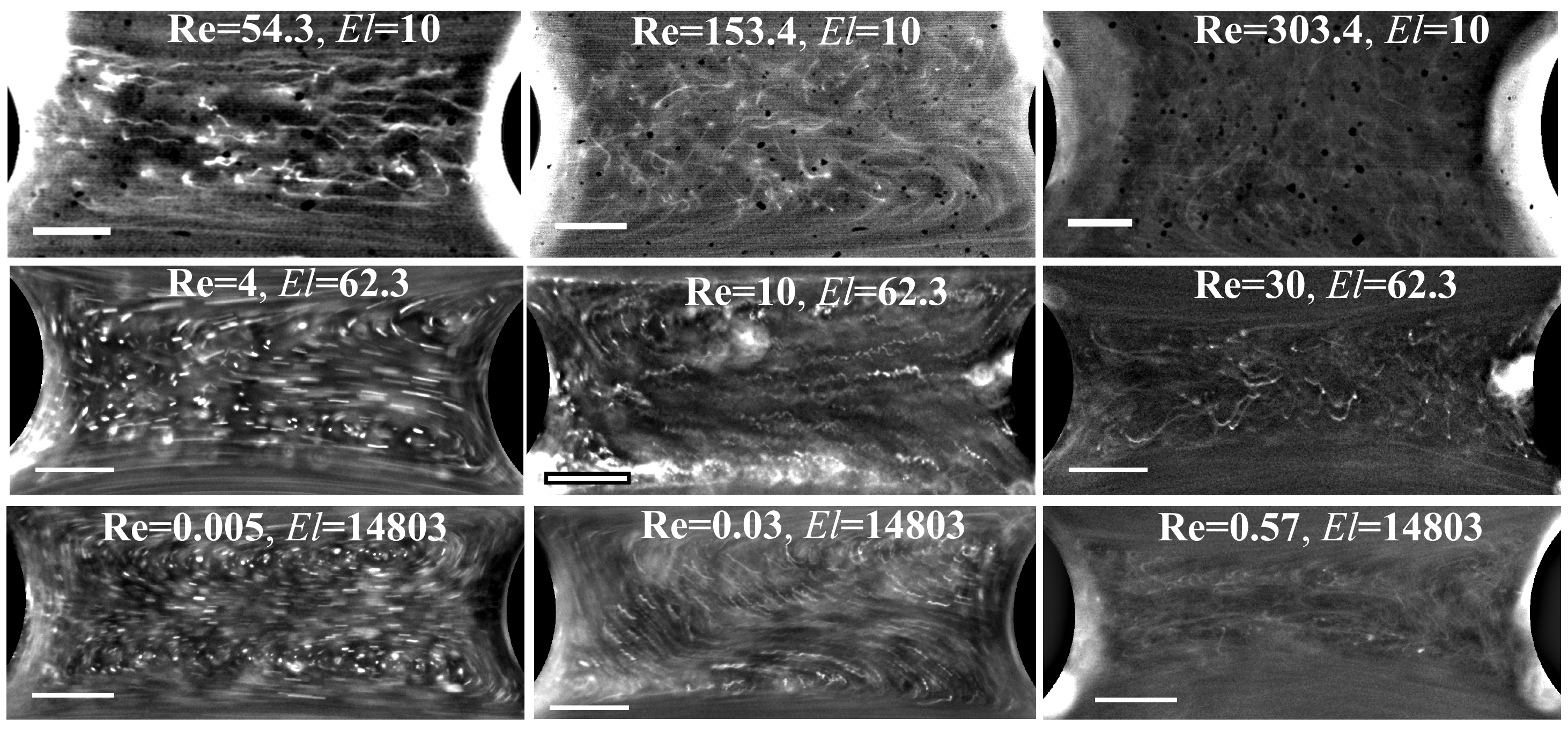}
		\caption{Representative snapshots of flow structures in the three regions, above the transitions, at $El=10$ (top panel), 62.3 (middle panel) and 14803 (bottom panel); see also corresponding SM1-SM9 Movies in \cite{sm}.   Scale bars, $100~\mu m$.}
\label{fig:streakimages}
	\end{center}
\end{figure*}

To elucidate further, the critical values of the respective transitions for each $El$ is mapped in $\mbox{Re}-El$ (Fig. \ref{fig:ReWivsEldiagram}(a)), $\mbox{Wi}-El$ (Fig. \ref{fig:ReWivsEldiagram}(b)) and $\mbox{Wi}-\mbox{Re}$ (Fig. \ref{fig:ReWivsEldiagram}(c)) coordinates. In the range explored for ($\mbox{Re}, \mbox{Wi}$) three different transitions are observed, which are associated with elastic instability, drag enhancement and DR as shown in Fig. \ref{fig:ReWivsEldiagram}(a,b). These transitions persist for all elasticity values and the elastic instability  occurs first followed by other two transitions. In addition, the complete flow relaminarization is observed only for $El=1070$, 2433 and 3704. Interestingly, the sequence of DR and drag enhancement changes as $El$ varies from low to high values;  DR is followed by drag enhancement at low $El$ and this sequence reverses at high $El$, as described above. This change in the sequence occurs in the intermediate range of elasticity at $El\sim149$. Furthermore, three regions in Fig. \ref{fig:ReWivsEldiagram}(a-c) can be identified based on variation of the critical values ($\mbox{Re}_{tr}, \mbox{Wi}_{tr}$) with $El$.  For low elasticity  ($El\leq20$), $\mbox{Re}_{tr}$ is independent of $El$, while for high elasticity ($El\geq300$), $\mbox{Re}_{tr}$ drops sharply with $El$. For intermediate elasticity ($20\leq El\leq300$), $\mbox{Re}_{tr}$ shows weak dependence on $El$ (see Fig. \ref{fig:ReWivsEldiagram}(a)). In Fig. \ref{fig:ReWivsEldiagram}(b), the dependence of $Wi_{tr}$ on $El$  is non-monotonic: a strong growth with $El$ at low $El$, a sharp decrease at high $El$, and a gradual growth at  intermediate $El$. The  transitions are further mapped in $\mbox{Wi}-\mbox{Re}$ plane for different $El$ to emphasize the role of inertia on the stability of a viscoelastic fluid flow.  Same three regions are identified: at high $El$,  $\mbox{Wi}_{tr}$ grows with $\mbox{Re}_{tr}$ with a stabilizing effect of inertia, at low $El$ a steep drop of $\mbox{Wi}_{tr}$ with $\mbox{Re}_{tr}$, and in the intermediate region $\mbox{Wi}_{tr}$ decreases with increasing $\mbox{Re}_{tr}$ with destabilizing effect of inertia (see Fig. \ref{fig:ReWivsEldiagram}(c)).
 \begin{figure*}
	\begin{center}
		\includegraphics[width=16cm]{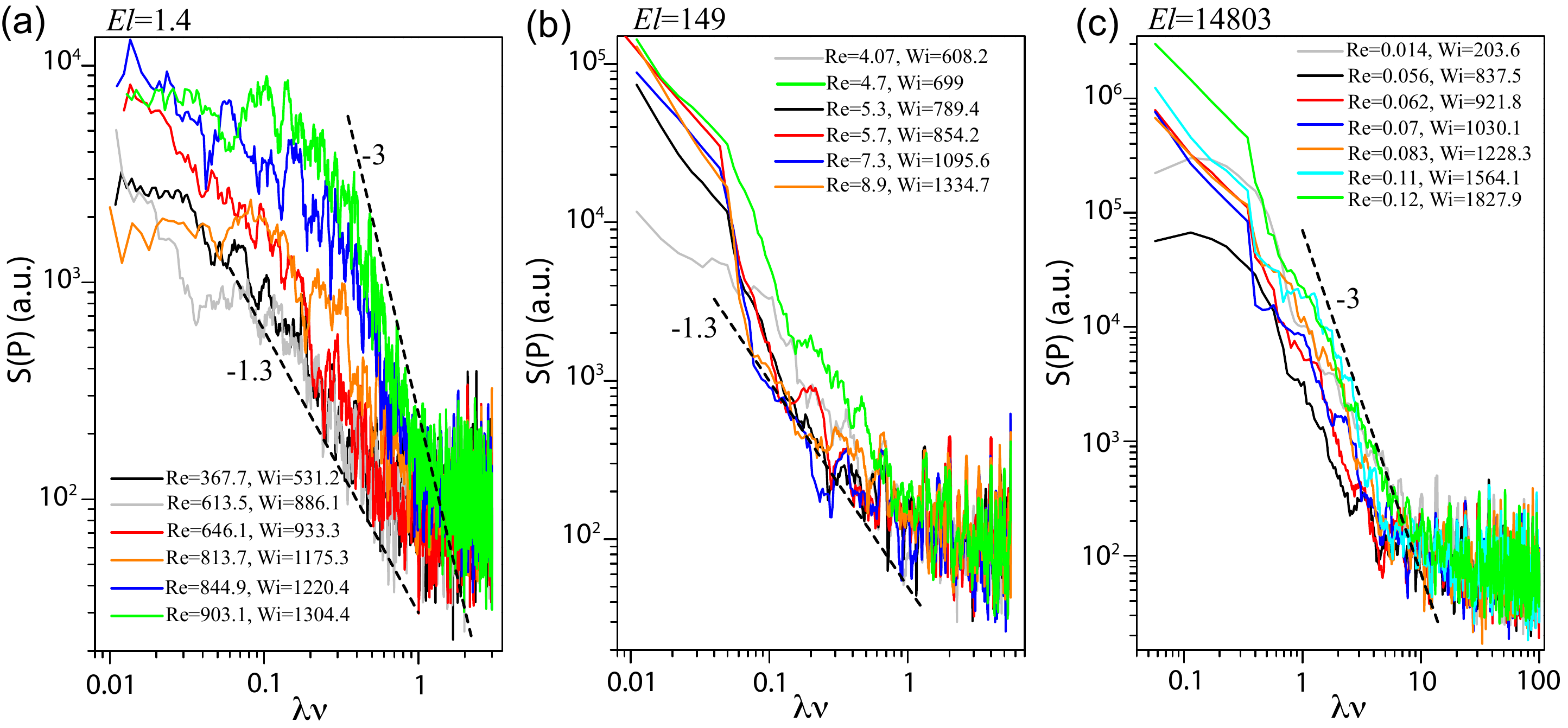}
		\caption{ Pressure power spectra $S(P)$ versus normalized frequency $\lambda\nu$ in the drag enhancement regime at various $\mbox{Re}$ and $\mbox{Wi}$, and in three regions of elasticity,  $El$= (a) 1.4, (b) 149 and (c) 14803. Dashed line shows power-law decay with an exponent $\beta$ specified beside to the line. }
		\label{fig:spectra}
	\end{center}
\end{figure*}
\begin{figure*}
	\begin{center}
		\includegraphics[width=16cm]{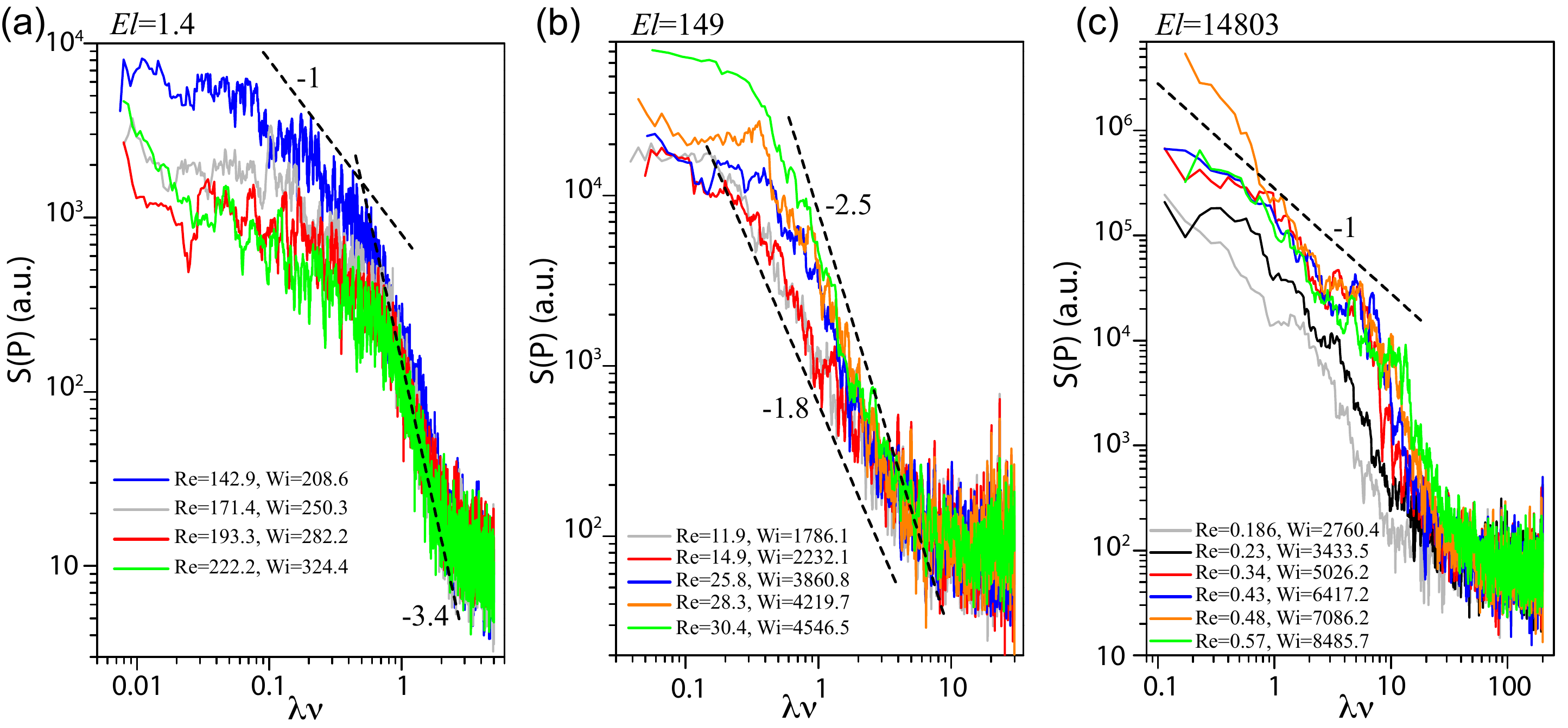}
		\caption{ Pressure power spectra $S(P)$ versus  $\lambda\nu$ in DR regime at various $\mbox{Re}$ and $\mbox{Wi}$, and in three regions of elasticity,  $El$= (a) 1.4, (b) 149 and (c) 14803. Dashed line shows power-law decay with an exponent specified beside to the line.}
		\label{fig:spectra1}
	\end{center}
\end{figure*}

Long-exposure particle streak images in Fig. \ref{fig:streakimages} illustrate the flow structures in three regions of elasticity and at different $\mbox{Re}$ and  $\mbox{Wi}$ above the transitions' values, see also corresponding S1-S9 Movies in \cite{sm}. In low and intermediate elasticity regions, a large scale vortical motion appears above the elastic instability, however, in DR and drag enhancement regimes small-scale turbulent structures dominate and  the large scale vortical motion vanishes (top and middle panels in Fig. \ref{fig:streakimages}). In high elasticity region, e.g. $El=14803$, unsteady pair of vortices \cite{atul} spans the region between the obstacles  (e.g. $\mbox{Re}=0.005,\mbox{Wi}=74$) and at higher $\mbox{Re}$  small scale vortices emerge with an intermittent and random dynamics (e.g. $\mbox{Re}=0.03,\mbox{Wi}=444$) that constitute the ET regime \cite{atul1}, whereas in DR regime (e.g. $\mbox{Re}=0.57,\mbox{Wi}=8438$) much smoother spatial scale and less vortical motion are found  (bottom panel of Fig. \ref{fig:streakimages}).  However, a quantitative analysis of velocity field  at low and intermediate values of $El$ requires serious technical efforts and will be the subject of our further investigation.

Finally, we characterize the observed flow regimes through frequency power spectrum of absolute pressure fluctuations  for various $\mbox{Re}$ and $\mbox{Wi}$ in three regions of elasticity. The  pressure spectra are presented as a function of normalized frequency $\lambda\nu$ to signify the time scales involved in flow  with respect to $\lambda$.  Figure \ref{fig:spectra} shows pressure power spectra $S(P)$ in the drag enhancement regime for three $El$ values. For low elasticity, $S(P)$ decay exponent $\beta$ evolves from -1.3 to -3 with increasing $\mbox{Re}$ and $\mbox{Wi}$ and in the range of $\lambda\nu\sim0.2$ to 1 (Fig. \ref{fig:spectra}(a) for $El=1.4$). It is worth to notice that $\beta\approx-3$ is reached at the highest $\mbox{Re}$ and $\mbox{Wi}$. In the intermediate range of elasticity, exponent value $\beta=-1.3$ is obtained in the range of $\lambda\nu\sim0.1$ to 1, same as for low $El$ (see  Fig. \ref{fig:spectra}(b) for $El=149$), whereas for high $El$, $S(P)$ exhibit steep decay with $\beta\approx-3$ in a higher frequency range $\lambda\nu\sim1$ to 10 for all $\mbox{Re}$ and $\mbox{Wi}$ values for $El=14803$ (Fig. \ref{fig:spectra}(c)). This value of $\beta$ is one of the main characteristics of the ET regime {\cite{jun,atul1}. The value $\lambda\nu=1$ at high $El$ is a relevant frequency to generate ET spectra with $\beta\approx-3$ at higher frequencies \cite{jun}, as stretching-and-folding mechanism of elastic stresses due to velocity field redistributes energy across the scales \cite{lebedev,lebedev1}. Similar scaling of $S(P)$ is observed in numerical simulations in the dissipation range of  the turbulent drag reduction regime \cite{gotoh}. For low $El$, $S(P)$ decay at $\lambda\nu>0.1$ up to 1 is caused by the inertial effect. In the drag reduction regime, $S(P)$ demonstrate completely different scaling behavior with $\lambda\nu$, shown in Fig. \ref{fig:spectra1}. For low $El$, one finds a steep decay of $S(P)$ at high frequencies $\lambda\nu\geq1$ with a scaling exponent $\sim-3.4$ and a rather slow decay with an exponent between -0.5 and -1 at $\lambda\nu<1$ (Fig. \ref{fig:spectra1}(a) for $El=1.4$), in accord with numerical simulations \cite{gotoh}. For high $El$, the spectra $S(P)$ decay steeply at  high frequency $\lambda\nu\sim10$, and at low frequencies $0.1<\lambda\nu\sim10$ a slow decay with an exponent $\sim-1$ is observed (Fig. \ref{fig:spectra1}(c) for $El=14803$). In the intermediate range of $El$, the decay exponent varies between -1.8 and -2.5 (Fig. \ref{fig:spectra1}(b) for $El=149$) in the frequency range $0.1<\lambda\nu\sim10$. To highlight the scaling dependencies of $S(P)$ between different flow regimes in each $El$ region, we present the same data in Fig. S1 in \cite{sm} for various $\mbox{Re}$ and $\mbox{Wi}$, and for three regions of $El$.

For comparison we present  $f$ as well as $f/f_{lam}$ as a function of $\mbox{Re}$ in the range between $\sim6$ and $\sim 900$ for two Newtonian fluids, water ($\eta_s=1~mPa\cdot s$) and a solution of $25\%$ sucrose in water ($\eta_s=3~mPa\cdot s$), see Fig. S2 in \cite{sm}. The dependencies of both $f$ and $f/f_{lam}$ on $\mbox{Re}$ are smooth and growing at $Re>\sim70$  that differs significantly from that found for polymer solutions at low $El$. It is also rather different from the dependence of $f$ on $Re$ in a channel flow past an obstacle, which is studied extensively. Thus, we can conclude that in viscoelastic flow, we observe   inertia-modified elastic instabilities, contrary to inertial instabilities modified by elastic stress.   This conclusion is further supported by the measurements of the power spectra of pressure fluctuations in a Newtonian fluid flow that exhibit power-law exponent $\sim-1.6$ at high $\mbox{Re}$ (see Fig. S3 in \cite{sm}), contrast to those presented in Figs. \ref{fig:spectra} and \ref{fig:spectra1}.

\section{Discussion and Conclusion}
Polymer degradation is often encountered under strong shear and in particular at high elongation rates due to velocity fluctuations at $\mbox{Re}\gg1$ and $\mbox{Wi}\gg1$ \cite{Gyr}. As a result of degradation, the influence of polymers on the flow becomes ineffective. To ensure that  drag enhancement and DR we observe in our experiments at $\mbox{Re}\geq1$ is not the result of polymer degradation, we reuse the polymer solutions (after performing the experiment with two obstacles) in experiments on a channel flow with a single obstacle. Indeed we observe elastic instability, drag enhancement and DR with single obstacle for $El=14803$, e.g. see Fig. S4 in \cite{sm} as an example and all results will be published elsewhere.   Moreover, the problem of polymer degradation was addressed in details in our paper on turbulent drag reduction in a large-scale swirling flow experiment conducted at $\mbox{Re} \leq2\times10^6$ \cite{burnishev}. It was pointed out that ``the main technical achievement in the experiment was long term stability of polymers in turbulent flow that allowed us to take large data sets up to $10^6$ data points for up to 3.5 hours at the highest $\mbox{Re}$ without a sign of polymer degradation.'' Thus, we conclude that the observed flow regimes in our experiments are not caused by polymer degradation.

The presented results on the friction coefficient and the pressure power spectra obtained in a wide range of the controlled parameters exhibit two remarkable features: (i) presence of three flow regimes with distinctive and different scaling behavior in both $f/f_{lam}$ and $S(P)$; and (ii) three regions on the stability diagrams in the planes of $\mbox{Re}$, $\mbox{Wi}$ and $El$ parameters depending on the value of fluid elasticity. In spite of the fact that rather high values of $\mbox{Re}$ are reached, inertial turbulence is not attained in the region between the obstacles and channel flow outside this region. As known from literature, turbulence in a flow past obstacle is attained at much higher $\mbox{Re}$ \cite{kundu}.

The different scaling dependencies of $S(P)$ in three flow regimes and in three regions of elasticity indicate  the intricate interaction between elastic and inertial stresses. A two-way of energy transfer between turbulent  kinetic energy and elastic energy of polymers also results in a modification of the velocity spectra scaling exponents  at $Re\gg1$   \cite{dubief,dubief1}. Effect of inertia at $\mbox{Re}\sim100$ on scaling behavior of velocity power spectra with the exponent $|\alpha|\approx2.2$ instead of $\sim3.5$ in pure ET   was first observed experimentally in the Couette-Taylor viscoelastic flow \cite{groisman2} and later confirmed numerically \cite{khomami}. What is remarkable that in the drag enhancement regime about the same scaling exponent $\beta\approx-3$ in $S(P)$ is found for low and high $El$ at close values of $\mbox{Wi}$, and three orders of magnitude difference in $\mbox{Re}$ values. It indicates the elastic nature of  drag enhancement regimes both at low and high $El$. Indeed, the scaling exponents of the pressure power spectra decay for $El=1.4$ (Fig. \ref{fig:spectra}(a)) show $|\beta|\cong3$ at $\mbox{Re}>\sim845$ and $\mbox{Wi}>\sim1220$. The observations of scaling $f/f_{lam}\sim Wi^{0.2}$,  exponent of pressure spectra decay $|\beta|\sim3$ and exponent of  velocity spectra decay $|\alpha|\sim3.5$ are characteristics of ET flow \cite{atul1}. Thus, the drag enhancement regime in low $El$ regions is typical of ET.

The striking and unanticipated observation, in high elasticity region,  is significant DR and a complete flow relaminarization at $Wi> 1000$ and $\mbox{Re}\sim\mathcal{O}(1)$ (see Figs. \ref{fig:dragcoeff} and \ref{fig:ReWivsEldiagram}). The obtained result is different from turbulent DR observed at $\mbox{Re}\gg1$, where Reynolds stress exceeds the elastic one prior to the onset of turbulent DR and becomes comparable  to elastic stress at the onset. Similar effect of the saturation and even weak reduction of $f/f_{lam}$ was observed and discussed in the planar geometry with an abrupt contraction-expansion of a micro-fluidic channel flow, where the saturation of $f/f_{lam}$ at higher polymer concentrations $c$ and even its reduction at lower $c<c^{*}$ were revealed in the range $0<Wi<500$ for three polymer solutions of different polymer concentrations \cite{mckinley}. For the highest $c$, $f/f_{lam}$ reached a value of $\sim3.5$ at high $Wi$ in agreement with the early measurements in a pipe flow with an axisymmetric contraction-expansion at much lower $Wi<8$ \cite{mckinley1}.

To find a possible explanation of DR in a wide range of $El$ and $(\mbox{Re},\mbox{Wi})$, we discuss the effect in details. At low $El$ between 1.4 and 31, either drag saturation or weak DR occurs just before the drag enhancement regime associated with ET and discussed above. It is worth to mention that due to the intricate interplay between elastic and inertial stresses the strength of DR is a non-monotonic function of $El$ and depends on the relation between $\mbox{Wi}$ and $\mbox{Re}$. The higher $Wi$ and the lower $Re$, the more pronounced DR regime at low $El$.  The range of $Re$ observed in DR regime corresponds to the vorticity suppression by elastic stress generated by polymer additives injected  into a Newtonian fluid flow \cite{cadot2,cadot3,goldburg}, which is indeed confirmed by the snapshots at $El=10$ and $\mbox{Re}=54$ and 153, shown in Fig. \ref{fig:streakimages}.

At high $El$ in the range $149<El<14803$ and $\mbox{Re}<\sim70$, $f/f_{lam}$ reduces significantly. However, the complete flow relaminarization is observed only at $El=1070$, 2433 and 3704, where $\mbox{Wi}\leq10^4$ and $\mbox{Re}<10$. It means that  high values of $\mbox{Wi}$ and $\mbox{Re}$ stabilize DR prior to the  relaminarization due to finite polymer extensibility. The snapshot at $El=14803$ and $\mbox{Re}=0.57$ in Fig. \ref{fig:streakimages} shows a vorticity-free flow, contrast to the snapshots at the same $El$ and low $\mbox{Re}$. Thus, at $\mbox{Re}\sim \mathcal{O}(1)$ and high $\mbox{Wi}$, the inertial effects are negligible to suppress the growth of $f/f_{lam}$, whereas at $\mbox{Re}\geq70$ and intermediate values of $\mbox{Wi}$, the drag can saturate, as seen, for example, at $El=31$.  As suggested in Ref. \cite{mckinley}, the saturation of $f/f_{lam}$ observed for high $El$  is probably a consequence of polymer chains attaining their finite extensibility limit at very high $\mbox{Wi}$ values. Here, we emphasize again that DR observed,  at low $El$ and high $\mbox{Re}$, is not related to {\bf turbulent drag reduction} realised in a turbulent flow  at  higher values of $\mbox{Re}$ than achieved in our experiment.

 The theory of ET and the corresponding numerical simulations do not consider the inertial effects and their role in ET,  and therefore they are unable to explain the DR and flow relaminarization.  Thus, the results reported call for further theoretical and numerical development to uncover inertial effects on viscoelastic flow in a broad range of ($\mbox{Re},\mbox{Wi}$) and $El$.

\section{Acknowledgements}

We thank Guy Han and Yuri Burnishev for technical support.  We are grateful to Dr. Dongyang Li for his help in the measurements of the friction factor and the pressure  spectra of Newtonian fluid flow. A.V. acknowledges support from the European Union's Horizon 2020 research and innovation programme under the Marie Sk{\l}odowska-Curie grant agreement No. 754411. This work was partially supported by the grants from Israel Science Foundation (ISF; grant \#882/15) and Binational USA-Israel Foundation (BSF; grant \#2016145).


\begin{thebibliography}{40}
	
	\bibitem{larson}
	R. G. Larson,
	\newblock Instabilities in viscoelastic flows.
	\newblock {\em Rheol. Acta} {\bf 31}, 213 (1992).
	
	\bibitem{shaqfeh}
	E. S. G. Shaqfeh,
	\newblock Purely elastic instabilities in viscometric flows.
	\newblock {\em Annu. Rev. Fluid Mech.} {\bf 28}, 129 (1996).
	
	\bibitem{groisman}
	A. Groisman \& V. Steinberg,
	\newblock Elastic turbulence in a polymer solution flow.
	\newblock {\em Nature} {\bf 405}, 53 (2000).
	
   \bibitem{bird}
	R. B. Bird, R. C. Armstrong \& O. Hassager,
	\newblock {\em Dynamics of polymeric liquids: Fluid mechanics}, Vol. 1 $\&$ 2.
	\newblock John Wiley $\&$ Sons., 2 edition  (1987).

	\bibitem{groisman1}
	A. Groisman \& V. Steinberg,
	\newblock Efficient mixing at low {Reynolds} numbers using polymer additives.
	\newblock {\em Nature} {\bf 410}, 905 (2001).
	
	\bibitem{groisman2}
	A. Groisman \& V. Steinberg,
	\newblock Elastic turbulence in curvilinear flows of polymer solutions.
	\newblock {\em New J. Phys.} {\bf 6}, 29 (2004).
	
	\bibitem{teo1}
	T. Burghelea, E. Segre \& V. Steinberg,
	\newblock Elastic turbulence in von {Karman} swirling flow between two disks.
	\newblock {\em Phys. Fluids} { \bf 19}, 053104 (2007).
	
	\bibitem{jun}
	Y. Jun \& V. Steinberg,
	\newblock Power and pressure fluctuations in elastic turbulence over a
	wide range of polymer concentrations.
	\newblock {\em Phys. Rev. Lett.}{ \bf 102}, 124503 (2009).
	
	\bibitem{jun1}
	Y. Jun \& V. Steinberg,
	\newblock Elastic turbulence in a curvilinear channel flow.
	\newblock {\em Phys. Rev. E}{ \bf 84}, 056325 (2011).
	
	\bibitem{arratia}
	L. Pan, A. Morozov, C. Wagner, P. E. Arratia,
	\newblock Nonlinear elastic instability and the transition to turbulence at low Reynolds numbers.
	\newblock {\em Phys. Rev. Lett.} {\bf 110}, 174502 (2013).
	
	\bibitem{atul1}
	A. Varshney \& V. Steinberg,
	\newblock Mixing layer instability and vorticity amplification in a creeping viscoelastic flow.
	\newblock {\em Phys. Rev. Fluids}, submitted (2018).
	
	\bibitem{lebedev}
	E. Balkovsky, A. Fouxon, V. Lebedev,
	\newblock Turbulence of polymer solutions.
	\newblock {\em Phys. Rev. E} {\bf 64}, 056301 (2001).
	
	\bibitem{lebedev1}
	A. Fouxon \& V. Lebedev,
	\newblock Spectra of turbulence in dilute polymer solutions.
	\newblock {\em Phys. Fluids} {\bf 15}, 2060 (2003).
	
	\bibitem{boffetta}
	G. Boffetta, A. Celani, \& S. Musacchio,
	\newblock Two-dimensional elastic turbulence in dilute polymer solutions.
	\newblock {\em Phys. Rev. Lett.} {\bf 91}, 034501 (2003).
	
	\bibitem{berti}
	S. Berti, A. Bistagnino, G. Boffetta, A. Celani \& S. Musacchio,
	\newblock Small scale statistics of viscoelastic turbulence.
	\newblock {\em EPL} {\bf 76}, 63 (2006).
	
	\bibitem{berti1}
	S. Berti, A. Bistagnino, G. Boffetta, A. Celani \& S. Musacchio,
	\newblock Two-dimensional elastic turbulence.
	\newblock {\em Phys. Rev. E} { \bf 77}, 055306(R) (2008).
	
	\bibitem{gotoh}
	T. Watanabe \& T. Gotoh,
	\newblock Power-law spectra formed by stretching polymers in decaying isotropic turbulence.
	\newblock {\em Phys. Fluids} {\bf 26}, 035110 (2014).
	
    \bibitem{toms}
	B. A. Toms,
	\newblock Some observations on the flow of linear polymer solutions through straight tubes at large Reynolds numbers.
	\newblock {in Proc. of the 1st International Congress of  Rheology} {Vol. 2}, {p. 135}, {North-Holland} (1949).

	\bibitem{hof}
	D. Samanta et al.,
	\newblock Elasto-inertial turbulence.
	\newblock {\em Proc. Natl. Acad. Sci. U.S.A.} {\bf 110}, 10557 {2013}.
	
	\bibitem{dubief}
	Y. Dubief, V. E. Terrapon \& J. Soria,
	\newblock On the mechanism of elasto-inertial turbulence.
	\newblock {\em Phys. Fluids} {\bf 25}, 110817 (2013).
	
	\bibitem{dubief1}
	V. E. Terrapon, Y. Dubief \& J. Soria,
	\newblock On the role of pressure in elasto-inertial turbulence.
	\newblock {\em J. Turbulence} {\bf 16}, 26 (2015).
	
	\bibitem{groismanprl96}
	A. Groisman \& V. Steinberg,
	\newblock Couette-Taylor flow in a  dilute polymer solution.
	\newblock {\em Phys. Rev. Lett.} {\bf 77}, 1480 (1996).
	
	\bibitem{groismanepl}
	A. Groisman \& V. Steinberg,
	\newblock Elastic {\it vs} inertial instability in a polymer solution flow.
	\newblock {\em EPL} {\bf 43}, 165 (1998).
	
   \bibitem{groismanphfl}
   A. Groisman \& V. Steinberg,
   \newblock Mechanism of elastic instability in Couette flow of polymer solutions: Experiment.
   \newblock {\em Phys. Fluids} {\bf 10}, 2451 (1998).

	\bibitem{shaqfeh1}
	Y. L. Joo \& E. S. G. Shaqfeh,
	\newblock The effects of inertia on the viscoelastic Dean and Taylor-Couette flow instabilities with application to coating flow.
	\newblock {\em Phys. Fluids A} {\bf 4}, 2415 (1992).
	
	\bibitem{thomas}
	D. G. Thomas, B. Khomami, R. Sureshkumar,
	\newblock Nonlinear dynamics of visco-elastic Taylo-Couette flow: effect of elasticity on pattern selection, molecular conformation and drag.
	\newblock {\em J. Fluid Mech.} {\bf 620}, 353 (2009).
	
   \bibitem{muller09}
   C. Dutcher \& S. Muller,
   \newblock The effects of drag reducing polymers on flow stability: Insights from the Taylor-Couette problem.
   \newblock {\em Korea-Australian Rheology Journal} {\bf 21}, 223 (2009).

   \bibitem{muller11}
   C. Dutcher \& S. Muller,
   \newblock Effects of weak elasticity on the stability of high Reynolds number co- and counter-rotating Taylor-Couette flows.
   \newblock {\em J. Rheol.} {\bf 55}, 1271 (2011).

   \bibitem{muller13}
   C. Dutcher \& S. Muller,
   \newblock Effects of moderate elasticity on the stability of co- and counter-rotating Taylor-Couette flows.
   \newblock {\em J. Rheol.} {\bf 57}, 791 (2013).

   \bibitem{suresh}
   B. Sadanandan \& R. Sureshkumar,
   \newblock Viscoelastic effects on the stability of wall-bounded shear flows.
   \newblock {\em Phys. Fluids} {\bf 14}, 41 (2002).

	\bibitem{mckinley}
	L. E. Rodd, T. P. Scott, D. V. Boger, J. J. Copper-White, G. H. McKinley,
	\newblock The inertio-elastic planar entry flow of low-viscosity elastic fluids in micro-fabricated geometries.
	\newblock {\em J. Non-Newtonian Fluid Mech.} {\bf 129}, 1 (2005).
	
	\bibitem{kenney}
	S. Kenney, K. Poper, G. Chapagain, \& G.-F. Christopher,
	\newblock Large {Deborah} number flows around confined microfluidic cylinders.
	\newblock {\em Rheol. Acta} {\bf 52}, 485 (2013).
	
	\bibitem{kellay}
	Y. L. Xiong, C. H. Bruneau \& H. Kellay,
	\newblock Drag enhancement and drag reduction in viscoelastic fluid flow around a cylinder.
	\newblock {\em EPL} {\bf 91}, 64001 (2010).
	
	\bibitem{liu3}
	Y. Liu, Y. Jun \& V. Steinberg,
	\newblock Concentration dependence of the longest relaxation times of dilute
	and semi-dilute polymer solutions.
	\newblock {\em J. Rheol.} {\bf 53}, 1069 (2009).
	
	\bibitem{atul}
	A. Varshney \& V. Steinberg,
	\newblock Elastic wake instabilities in a creeping flow between two obstacles.
	\newblock {\em Phys. Rev. Fluids}{ \bf 2}, 051301(R) (2017).
	
	
	\bibitem{sm}
    {See Supplemental Material for movies, figures and table}.

    \bibitem{Gyr}
	A. Gyr \& H.W. Bewersdorf,
	\newblock Drag Reduction of Turbulent Flows by Additives.
	\newblock Kluwer Academic Publishers, Dordrecht, (2003).
	
	
   \bibitem{burnishev}
   Y. Burnishev \& V. Steinberg,
   \newblock Influence of polymer additives on turbulence in von karman swirling flow between two disks, II.
   \newblock {\em Phys. Fluids} {\bf 28}, 033101 (2016).

   \bibitem{kundu}
   P. K. Kundu \& I. M. Cohen,
   \newblock {\em Fluid Mechanics}
   \newblock Elsevier, New York, $4^{th}$ ed., (2008).
	
\bibitem{khomami}
	N. Liu \& B. Khomami,
	\newblock Elasticity induced turbulence in Taylor-Couette flow: direct numerical  simulation and mechanistic insight.
	\newblock {\em J. Fluid Mech.} {\bf 737}, R4 (2013).

\bibitem{mckinley1}
	J. Rothstein \& G. H. McKinley,
	\newblock The axisymmetric contraction-expansion: the role of extensional rheology on vortex growth dynamics and the enhanced pressure drop.
	\newblock {\em J. Non-Newtonian Fluid Mech.} {\bf 98}, 33 (2001).

\bibitem{cadot2}{O. Cadot and M. Lebey, Shear instability inhibition in a cylinder wake by local injection of
  a viscoelastic fluid. {\sl Phys. Fluids} {\bf 11}, 494 (1999).}
\bibitem{cadot3}{O. Cadot and S. Kumar, Experimental characterization of viscoelastic effects on two- and three-dimensional shear instabilities. {\sl J. Fluid Mech. } {\bf 416}, 151 (2000).}
\bibitem{goldburg}{J. R. Cressman, Q. Baley, and W. I. Goldburg, Modification of a vortex street by a polymer additive. {\sl Phys. Fluids} {\bf 13}, 867 (2001).}
\end{thebibliography}
\end{document}